\documentclass[rnote]{aa}
\newcommand{\GILDAS}{\texttt{GILDAS}}
\newcommand{\CLASS}{\texttt{CLASS}}
\usepackage{natbib}
\usepackage{amsmath}
\usepackage{graphicx}
\usepackage{txfonts}
\usepackage{subfig}
\usepackage{multirow} 
\usepackage{threeparttable}
\usepackage{color}
\newcommand{\ie}{\emph{i.e.}}
\newcommand{\eg}{e.g.}
\newcommand{\emm}[1]{\ensuremath{#1}}   
\newcommand{\emr}[1]{\emm{\mathrm{#1}}} 

\newcommand{\Cp}{\emr{C^{+}}}                  
\newcommand{\thCO}{\emr{^{13}CO}}                  
\newcommand{\CFp}{\emr{CF^{+}}}                    

\newcommand{\pscm}{~\rm{cm}^{-2}}

\newcommand{\kms}{~\rm{km}~\mathrm{s}^{-1}}

\begin{document}
\title{The hyperfine structure in the rotational spectrum of $\CFp$}

\authorrunning{V. Guzm\'an et al.}
\titlerunning{The hyperfine structure of $\CFp$}

\author{V. Guzm\'{a}n\inst{1} \and E. Roueff\inst{2} \and J. Gauss\inst{3} 
  \and J. Pety\inst{1,4} \and P. Gratier\inst{1} \and J.R.
  Goicoechea\inst{5} \and M. Gerin\inst{4} \and D. Teyssier\inst{6}}

          \institute{
            IRAM, 300 rue de la Piscine, 38406 Saint Martin d'H\`eres, France\\
            \email{guzman@iram.fr}
            \and
            LUTH UMR 8102, CNRS and Observatoire de Paris, Place J. Janssen, 
            92195 Meudon Cedex, France.
            \and
            Institut f\"ur Physikalische Chemie, Universit\"at Mainz, D-55099 Mainz, Germany
            \and
            LERMA - LRA, UMR 8112, Observatoire de Paris and Ecole 
            normale Sup\'{e}rieure, 24 rue Lhomond, 75231 Paris, France. 
            \and
            Centro de Astrobiolog\'{i}a. 
            CSIC-INTA. Carretera de Ajalvir, Km 4. Torrej\'{o}n de Ardoz, 
            28850 Madrid, Spain.
            \and
            European Space Astronomy Centre, ESA, PO Box 78, 28691
            Villanueva de la Ca\~nada, Madrid, Spain 
          }


\abstract %
{\CFp{} has recently been detected in the Horsehead and
Orion Bar photo-dissociation regions. The $J=1-0$ line in the
Horsehead is double-peaked in contrast to other millimeter lines. The
origin of this double-peak profile may be kinematic or spectroscopic.}
{We investigate the effect of hyperfine interactions due to the fluorine
  nucleus in \CFp{} on the rotational transitions.}
{We compute the fluorine spin rotation constant of \CFp{} using high-level
  quantum chemical methods and determine the relative positions and
  intensities of each hyperfine component. This information is used to
  fit the theoretical hyperfine components to the observed \CFp{} line
  profiles, thereby employing the hyperfine fitting method in \GILDAS{}.}
{The fluorine spin rotation constant of \CFp{} is 229.2~kHz. This way,
  the double-peaked \CFp{} line profiles are well fitted by the
  hyperfine components predicted by the calculations. The unusually
  large hyperfine splitting of the \CFp{} line therefore explains the
  shape of the lines detected in the Horsehead nebula, without
  invoking intricate kinematics in the UV-illuminated gas.}
{}
\maketitle
%

\newcommand{\TabShifts}{%
  \begin{table}
    \begin{center}
    \caption{\small Hyperfine splitting shifts with respect to the strongest
      transition and relative intensities.}
    \label{tab:hfs_shifts}
      {\tiny
        \begin{tabular}{ccrrc}
          \hline \hline
          Transition & Component  & Shift Freq. & Shift Vel. & Rel. intensity\\
          $J \rightarrow J-1$ & $F \rightarrow F^{'}$ & (kHz) & ($\kms$) & \%\\
          \hline
          \multirow{2}{*}{$1-0$} & $3/2-1/2$ & 0      & 0.00 & 66.7\\
                                 & $1/2-1/2$ & -343.8 & 1.00 & 33.3\\
          \hline
          \multirow{3}{*}{$2-1$} & $5/2-3/2$ & 0.00   & 0.00 & 60.0\\
                                 & $3/2-1/2$ & -229.2 & 0.33 & 33.4\\
                                 & $3/2-3/2$ & -573.0 & 0.84 & 6.66\\
          \hline
        \end{tabular}}
    \end{center}
\end{table}}

\newcommand{\TabHFS}{%
  \begin{table}
    \begin{center}
    \caption{\small Results of the Hyperfine component fit.}
    \label{tab:hfs_fit}
      {\tiny
        \begin{tabular}{llcccccccr}
          \hline \hline Line & $v_{\textrm{LSR}}$ & Width &
          $\tau_{\textrm{main}}$\\ 
          & K & $\kms$ & \\ 
          \hline 
          $J=1-0$ & $10.70\pm0.01$ & $0.65\pm0.02$ & $0.10\pm0.96$\\ 
          $J=2-1$ & $10.70\pm0.02$ & $0.49\pm0.05$ & $0.10\pm0.45$\\ 
          \hline
        \end{tabular}}
    \end{center}
\end{table}
}

\newcommand{\FigLines}{%
  \begin{figure}[t!]
    \centering %
    \includegraphics[width=0.39\textwidth{}]{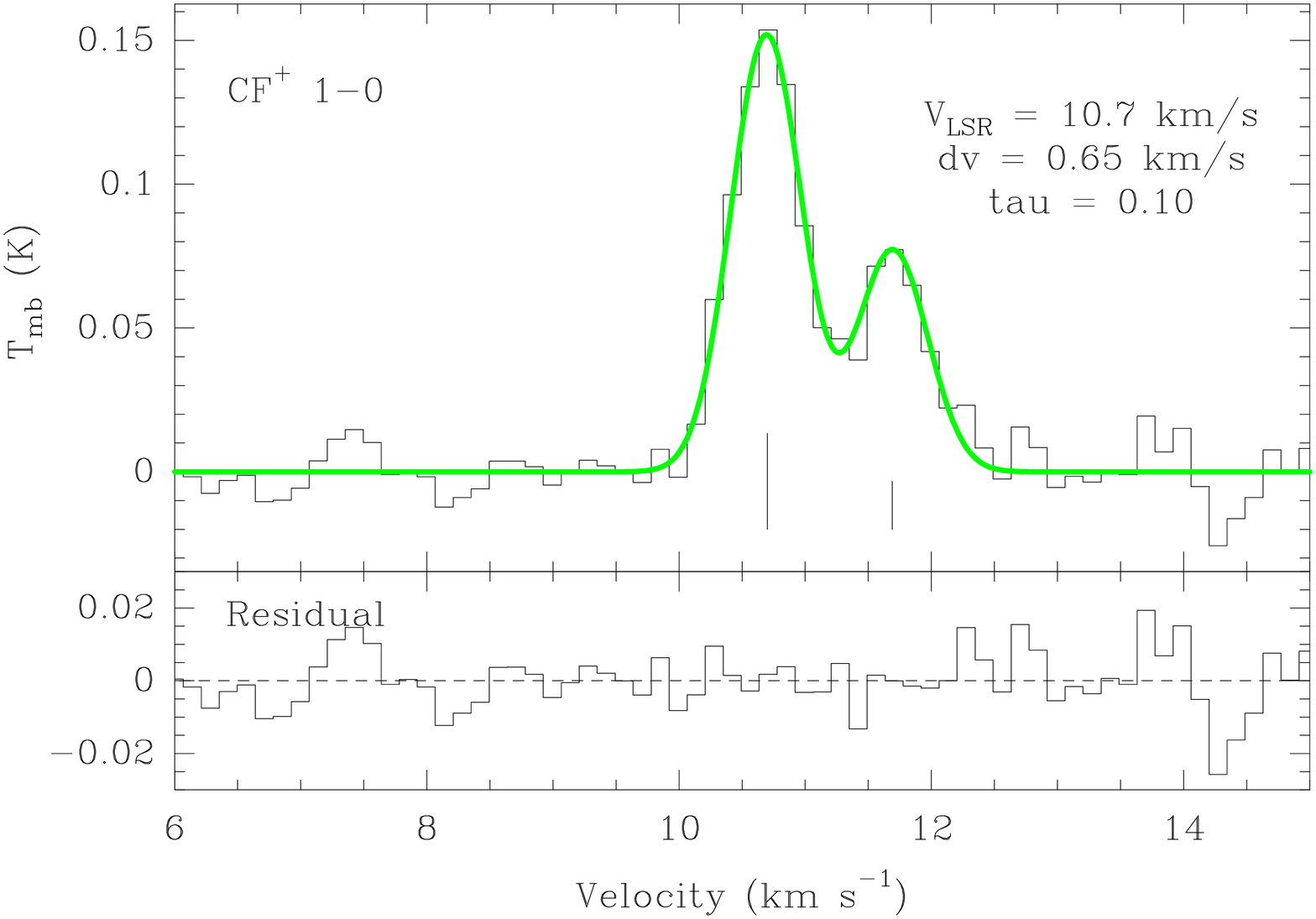}\\ 
    \includegraphics[width=0.39\textwidth{}]{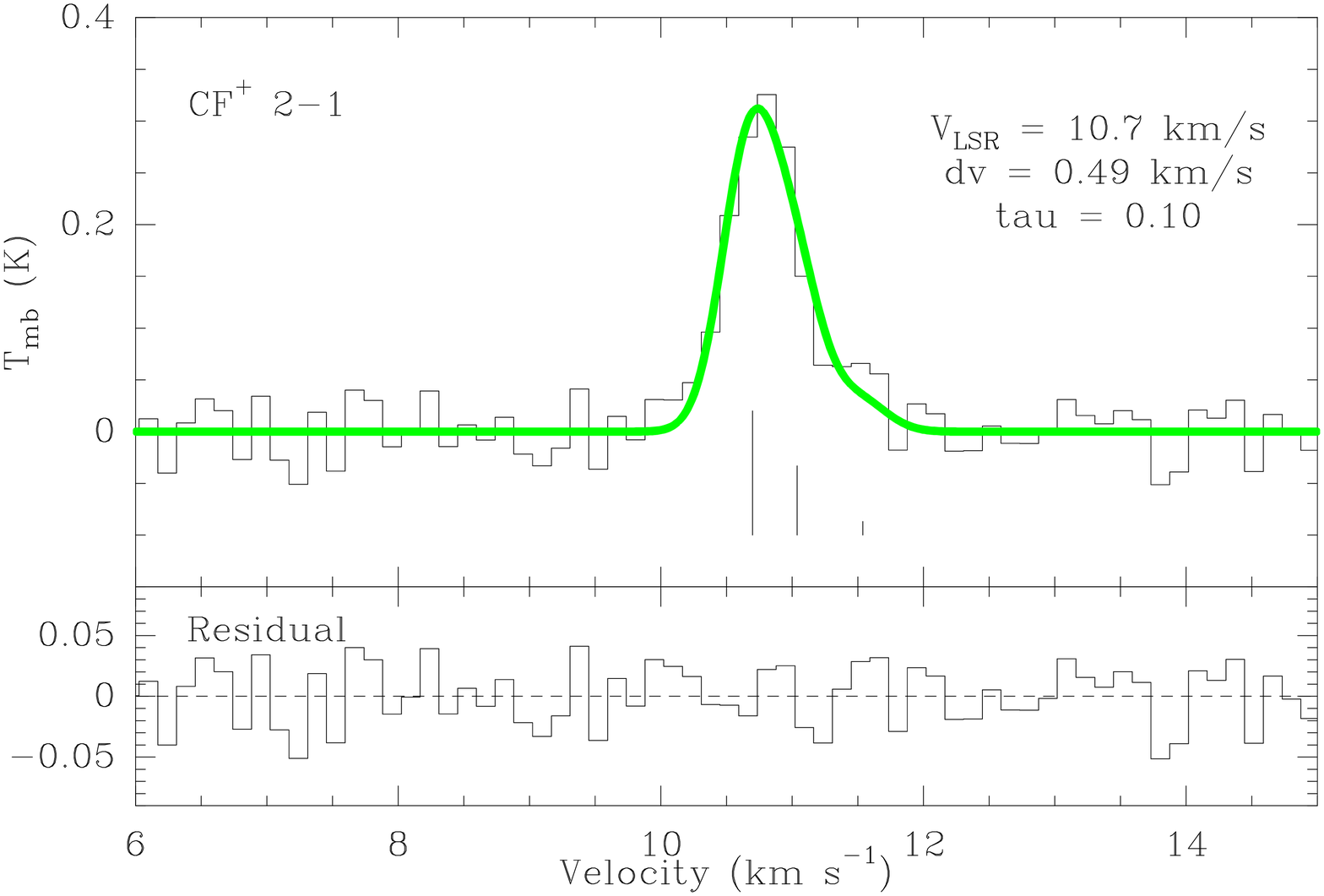}
    \caption{\small $\CFp$ $J=1-0$ and $J=2-1$ line profiles detected toward
      the Horsehead PDR. The vertical lines indicate the positions and the
      relative intensities of the hyperfine components. The green lines
      display the hyperfine fit performed with the CLASS tool in \GILDAS{}.
      The fit residuals are shown in the bottom panels.}
    \label{fig:lines}
  \end{figure}
}


\section{Introduction}

\vspace*{-0.15cm}

Up to now, \CFp{} has only been detected in two photo-dissociation
regions (PDR), namely the Horsehead mane~\citep{guzman12} and the
Orion bar~\citep{neufeld06}. In contrast to other mm lines in the
Horsehead, the \CFp{} $J=1-0$ (102.587~GHz) line is
double-peaked. \citet{guzman12} proposed three possible explanations:
1) The \CFp{} line suffers from self-absorption, 2) the complex \CFp{}
line profile is caused by kinematics, or 3) the two peaks correspond
to different hyperfine components caused by the fluorine nuclei. As
the \CFp{} opacities are low ($\tau \la 1$), the radiative transfer
explanation could be ruled out.  However, \citet{guzman12} could not
distinguish between the remaining possibilities for two
reasons. First, \CFp{} is expected to coexist with \Cp{}, because
\CFp{} is formed by the reaction of HF with \Cp{}. This implies that
the lines of both \Cp{} and \CFp{} may share similar kinematics and
indeed the 157.8~$\mu$m \Cp{} line is double-peaked. Second, no
experimental or theoretical investigation of the hyperfine structure
of \CFp{} was available. The most recent spectroscopic study of \CFp{}
did not consider the hyperfine structure because the experiments were
not able to resolve the hyperfine
components~\citep{cazzoli10}. High-level quantum chemical calculations
reported in the following now allow us to determine the hyperfine
splitting. It is thus possible to establish the spectroscopic origin
of the double-peak line in the Horsehead PDR.

\vspace*{-0.5cm}
\section{Hyperfine structure}
\vspace*{-0.15cm}

\FigLines{}

$\CFp$ has a $^1\Sigma^+$ electronic ground state with a dipole moment
of 1.04~Debye \citep{peterson1990}. The fluorine nuclei has a non-zero
nuclear spin ($I=1/2$). Therefore, the rotational and spin angular
momenta are coupled through $\vec{J} + \vec{I} = \vec{F}$. This
implies that the total angular momentum $F$ can have the values
$F=I+J,I+J-1,...,|I-J|$, where $J$ is the rotational and $I$ the
nuclear spin quantum number.  The various combinations of these
quantum numbers are responsible for the hyperfine splitting of the
energy levels, which are given by
\begin{equation}
E = E_J + \frac{C_I}{2} \left [ F(F+1) - I(I+1) - J(J+1) \right ],
\end{equation}
with $C_I$ as the spin rotation constant. The frequency splitting of
the two hyperfine lines is equal to $3/2~C_I$ for the $J=1-0$
transition. For the $J=2-1$ transition, there are three hyperfine
lines separated by $C_I$ and $5/2~C_I$ from the strongest
transition~\citep{townes75}.

The \CFp{} spin rotation constant has been determined by means of
quantum chemical calculations~\citep[see, \eg{},][for
  details]{puzzarini2010}. To ensure high accuracy, the computations
were performed at the coupled-cluster singles and doubles level
augmented by a perturbative treatment of triple
excitations~\citep[CCSD(T),][]{raghavachari1989} together with
atomic-orbital basis sets from Dunning's cc-pCVXZ hierarchy with X =
Q, 5, and 6~\citep{woon1995}.  Rotational London orbitals
\citep{gauss1996} were used in all calculations.  Furthermore,
zero-point vibrational effects were accounted for in a perturbative
manner as described in \citet{Auer2003} at the CCSD(T)/cc-pCVQZ level.
The values computed for $r(\emr{CF})=1.154089\AA$ are 227.5 kHz
(cc-pCVQZ), 228.5 kHz (cc-pCV5Z), and 228.7 kHz (cc-pCV6Z),
respectively. Vibrational corrections amount to 0.55 kHz, thus leading
to a final theoretical value of 229.2~kHz for the spin rotation
constant of \CFp.  All calculations were performed with the quantum
chemical program package {\sc CFour}\footnote{See
  \texttt{http://www.cfour.de} for more information about the {\sc
    CFour} package.}.

Table~\ref{tab:hfs_shifts} gives the resulting hyperfine frequency
shifts together with the relative line intensities, taking a \CFp{}
spin rotation constant of 229.2~kHz. This value is higher than, \eg{},
the \thCO{} value of 32.6~kHz~\citep{cazzoli04}. This makes \CFp{} a
spectroscopically peculiar ion. Indeed, extrapolating from the
spectroscopy of $^{13}$CO, \citet{guzman12} significantly
underestimated the hyperfine splitting of the $J=1-0$ line by assuming
a value of about 165~kHz.  The computed value of the spin rotation
constant agrees with the fact that the hyperfine splittings were not
observed in the experiments of \cite{cazzoli10}.

\vspace*{-0.3cm}
\section{Comparison with observations}

The hyperfine fitting method available in the \GILDAS{}/\CLASS{}
software\footnote{See \texttt{http://www.iram.fr/IRAMFR/GILDAS} for
  more information about the \GILDAS{} softwares.} was used to fit the
\CFp{} hyperfine components to the available astronomical data. The
fits and their residuals are shown in Figure~1. Table~2 gives the
results of the fits, namely the local standard of rest (LSR) velocity,
the linewidth, and the total opacity of the line. The observed lines
are well fitted and the column density inferred from the fitted total
opacity is $N=(1.8-3.5)\times10^{12}\pscm$, \ie{}, less than a factor
two from the value inferred in~\citet{guzman12}. The frequency
splitting and relative intensities of the two peaks in the
$\CFp~J=1-0$ line emission profile are consistent with two hyperfine
components due to the fluorine nucleus. The two strongest hyperfine
components in the $J=2-1$ line are separated by only $0.3\kms$ at the
LSR velocity associated to the Horsehead nebula while the typical
linewidth of the lines is $\sim0.6\kms$. The velocity separation of
the third hyperfine component is much larger, but its relative
intensity is weak (less than 10\%). Therefore, the three hyperfine
components are not easily disentangled at 1~mm.  Nevertheless, the
line profile is also compatible with the computed hyperfine splitting,
as shown on the residuals. The fitted velocities for both
lines are $10.7\kms$; i.e., they are equal to the usual systematic
velocity of all the lines detected in the Horsehead WHISPER survey
(PI: J.~Pety). This way, the observed \CFp{} double-peak profile can
be fully explained spectroscopically. This result demonstrates the
importance of multidisciplinary collaborations to interpret
astrophysical observations.

 \TabShifts{} \TabHFS{} 
 
The difference in typical
linewidth explains why the hyperfine splitting was not detected in the
Orion bar \citep[$1.5-3.0\kms$,][]{neufeld06}. The Horsehead (showing
very narrow emission lines) is a perfect laboratory for precise
spectroscopic studies of species present in UV-illuminated
environments.

\begin{acknowledgements}
  VG acknowledges support from the Chilean Government through the Becas
  Chile scholarship program. This work was also funded by grant
  ANR-09-BLAN-0231-01 from the French {\it Agence Nationale de la
    Recherche} as part of the SCHISM project and in Mainz by the Deutsche
  Forschungsgemeinschaft through grant GA 370/5-1. JRG thanks the Spanish
  MICINN for support through grants AYA2009-07304 and CSD2009-00038. JRG is
  also supported by a Ram\'on y Cajal research contract from the Spanish
  MICINN and co-financed by the European Social Fund.
\end{acknowledgements}

\vspace*{-0.5cm}
\bibliographystyle{aa} 
\vspace{-0.2cm}
\bibliography{draft-cfp}

\end{document}